\documentclass[aps,prd,superscriptaddress,showpacs]{revtex4}

\usepackage{graphicx,epsfig}

\begin{document}

\title{Some properties of evolving wormhole geometries within nonlinear
electrodynamics}
\author{Aar\'{o}n V. B. Arellano}
\affiliation{Facultad de Ciencias, Universidad Aut\'onoma del Estado de
M\'exico, El Cerrillo, Piedras Blancas, 50200, Toluca, MEXICO}
\author{Nora Bret\'on }
\affiliation{Depto. de F\'{\i}sica, Centro de Investigaci\'on y de Estudios
Avanzados del I.P.N., Apdo. Postal 14-740, 07000, M\'exico, D.F., MEXICO}
\author{Ricardo Garc\'{\i}a-Salcedo}
\affiliation{Depto. de F\'{\i}sica Educativa, CICATA-IPN, Legaria 694, Del.
Miguel Hidalgo, 11500, M\'exico, D.F., MEXICO}

\begin{abstract}
In this paper we review some properties for the evolving wormhole solution
of Einstein equations coupled with nonlinear electrodynamics. We integrate
the geodesic equations in the effective geometry obeyed by photons; we check
out the weak field limit and find the traversability conditions. Then we
analyze the case when the lagrangian depends on two electromagnetic
invariants and it turns out that there is not a more general solution within
the assumed geometry.
\end{abstract}

\maketitle

\section{Introduction}
 
Recently, the interest in wormholes has increased because of the possibility
of interstellar travel or future time travel to past world. Since the
formulation of Einstein's equations, several features of these solutions were
addressed, such as Einstein-Rosen bridges \cite{ER}; Wheeler \cite{Wheeler}
analyzed this kind of solutions and coined the name of wormholes. The most
important and popular contribution to these solutions was given by Morris and
Thorne \cite{Morris}, they were the first who made a complete and detailed
investigation in this field. Wormholes are solutions to Einstein's field
equations which represent two connected universes or a connection between two
distant regions of universe. For realistic models of traversable wormholes,
one should address additional engineering issues such as tidal effects.
People is skeptic about them because they contradict some commonly accepted
reasonable energy conditions, for example, their structure requires matter
with negative energy density as a source, i. e. exotic matter \cite{Morris}.
A full revision on wormholes can be consulted in \cite{Visser} and
\cite{Lobo}.
 
Among an enormous variety of wormhole solutions, there are some kind of
spherically symmetric wormholes that evolve with time. This type of wormholes
have been extensively analyzed. Roman \cite{Roman} explored the possibility
that inflation might provide a mechanism to enlarge tiny wormholes to
macroscopic size and investigated the possibility of avoiding the violation
of the energy conditions in the process. Further evolving dynamic wormhole
geometries were analyzed, considering specific cases \cite{Kar1}-
\cite{Trobo}; and also was considered the evolution of a wormhole model in an
FRW background \cite{Kim-evolvWH}.

On the other hand, nonlinear electrodynamics has recently been applied in
several branches in physics, namely, as effective theories at different
levels of string/M-theory \cite{Witten}, cosmological models \cite{CM}, black
holes \cite{BH}- \cite{Bro1} and in wormhole physics \cite{WH1}, among
others. In this context, in a recent paper it was shown that (2 + 1) and (3 +
1)-dimensional static spherically symmetric as well as stationary
axisymmetric traversable wormholes cannot be supported by nonlinear
electrodynamics \cite{AL1}. However, it was found an evolving wormhole
solution within nonlinear electrodynamics \cite{AL2}, where the lagrangian
depends in nonlinear way on one electromagnetic invariant.

In this paper we explore some properties of the evolving wormhole geometry in
the context of nonlinear electrodynamics. This paper is outlined in the
following manner. In section II, we describe briefly (3+1)-dimensional
evolving spherically symmetric wormholes coupled with nonlinear
electrodynamics. Section III is devoted to the study of wormhole geometry in
the aspects of: traversability conditions, linear limit and photon
trajectories. In Section IV we analyze the case when the lagrangian depends
on the two electromagnetic invariants, attempting to find a more general
solution. In section V some conclusions are given. We shall use geometrized
units, i.e., G = c = 1, throughout this work.

\section{$(3+1)-$dimensional evolving wormhole solution}

\subsection{Action and spacetime metric}

The action of $(3+1)$-dimensional general relativity coupled to nonlinear
electrodynamics is given by

\begin{equation}
S=\int \sqrt{-g}\left[ \frac{R}{16\pi }+L(F)\right] \,d^{4}x\,,  \label{SF}
\end{equation}
where $R$ is the Ricci scalar. $L(F)$ is the lagrangian, depending in
nonlinear form on a single invariant $F$ given by $F=F^{\mu \nu }F_{\mu \nu
}/4$ \cite{Pleb}, where $F_{\mu \nu }$ is the electromagnetic tensor. Note
that in Einstein-Maxwell theory the lagrangian takes the form $L(F)=-F/4\pi$,
however, we shall consider more general electromagnetic lagrangians.

Varying the action with respect to the gravitational field provides the
Einstein field equations $G_{\mu\nu}=8\pi T_{\mu\nu}$, with the
stress-energy tensor given by

\begin{equation}
T_{\mu\nu}=g_{\mu\nu}\,L(F)-F_{\mu\alpha}F_{\nu}{}^{\alpha}\,L_{F}\,,
\label{4dim-stress-energy}
\end{equation}
where $L_F=dL/dF$. The variation with respect to the electromagnetic
potential $A_\mu$, where $F_{\mu\nu}=A_{\mu,\nu}-A_{\nu,\mu}$, yields the
electromagnetic field equations 
\begin{equation}
\left(F^{\mu\nu}\,L_{F}\right)_{;\mu}=0 \,.  \label{em-field}
\end{equation}

The spacetime metric representing a dynamic spherically symmetric $(3+1)$%
-dimensional wormhole, which is conformally related to the static wormhole
geometry \cite{Morris}, takes the form

\begin{equation}
ds^{2}=\Omega ^{2}(t)\left[ -e^{2\Phi (r)}dt^{2}+\frac{dr^{2}}{1-b(r)/r}
+r^{2}(d\theta ^{2}+\sin ^{2}{\theta }d\phi ^{2})\right] \,,  \label{4dysme}
\end{equation}
where $\Phi $ and $b$ are functions of $r$, and $\Omega =\Omega (t)$ is the
conformal factor that is finite and positive definite throughout the domain
of $t$. $\Phi $ is the redshift function, and $b(r)$ is the shape function 
\cite{Morris}. We shall also assume that these functions satisfy all the
conditions required for a wormhole solution, namely, $\Phi (r)$ is finite
everywhere in order to avoid the presence of event horizons; $b(r)/r<1$,
with $b(r_{0})=r_{0}$ at the throat; as well as the fulfilment of the
flaring out condition, $b-b^{\prime 2}\geq 0$, with $b^{\prime }(r_{0})<1$
at the throat.

Now, taking into account the metric (\ref{4dysme}) the electromagnetic
tensor compatible with the assumed symmetries is given by

\begin{equation}
F_{\mu \nu }=E(x^{\alpha })\left( \delta _{\quad \mu }^{t}\delta _{\quad \nu
}^{r}-\delta _{\quad \mu }^{r}\delta _{\quad \nu }^{t}\right) +B(x^{\alpha
})\left( \delta _{\quad \mu }^{\theta }\delta _{\quad \nu }^{\varphi
}-\delta _{\quad \mu }^{\varphi }\delta _{\quad \nu }^{\theta }\right) ,
\label{4em-tensor}
\end{equation}%
being the non-zero components the following: $F_{tr}=-F_{rt}=E$, the
electric field; and $F_{\theta \phi }=-F_{\phi \theta }=B$, the magnetic
field. The invariant $F$ takes the form 
\begin{equation}
F=-\frac{1}{2\Omega ^{4}}\left[ \left( 1-\frac{b}{r}\right) e^{-2\Phi
}\,E^{2}-\frac{B^{2}}{r^{4}\sin ^{2}\theta }\right] \,.  \label{invF}
\end{equation}

\subsection{Einstein field equations}

We recall that from \cite{AL2}, the solution requirement $\Phi =0$ 
must be imposed, so using an orthonormal reference frame the non zero
components of the Einstein tensor reduce to

\begin{eqnarray}
G_{\hat{t}\hat{t}} &=&\frac{1}{\Omega ^{2}}\left[ \frac{b^{\prime }}{r^{2}}
+3\left( \frac{\dot{\Omega}}{\Omega }\right) ^{2}\right] ,  \nonumber \\
G_{\hat{r}\hat{r}} &=&\frac{1}{\Omega ^{2}}\left\{ -\frac{b}{r^{3}}+\left[
\left( \frac{\dot{\Omega}}{\Omega }\right) ^{2}-2\frac{\ddot{\Omega}}{\Omega 
}\right] \right\} ,  \label{ET} \\
G_{\hat{\theta}\hat{\theta}} &=&G_{\hat{\varphi}\hat{\varphi}}=\frac{1}{
\Omega ^{2}}\left\{ -\frac{b^{\prime }r-b}{2r^{3}}+\left[ \left( \frac{\dot{
\Omega}}{\Omega }\right) ^{2}-2\frac{\ddot{\Omega}}{\Omega }\right] \right\},
\nonumber
\end{eqnarray}
where overdot denotes a derivative with respect to the time coordinate, $t,$
and the prime a derivative with respect to $r.$Analogously, the nonzero
components of the stress energy tensor, from Eq. (\ref{4dim-stress-energy}),
take the form

\begin{eqnarray}  \label{4rset}
T_{\hat{t}\hat{t}} &=&-L-\frac{(1-b/r)}{\Omega ^{4}}E^{2}L_{F}\,,
\label{4rTott} \\
T_{\hat{r}\hat{r}} &=&L+\frac{(1-b/r)}{\Omega ^{4}}E^{2}L_{F}\,,
\label{4rTorr} \\
T_{\hat{\theta}\hat{\theta}} &=&T_{\hat{\phi}\hat{\phi}}=L-\frac{1}{\Omega
^{4}r^{4}\sin ^{2}\theta }B^{2}L_{F}\,.  \label{4rTohhpp}
\end{eqnarray}

It is clear that $T_{\hat{t}\hat{t}}=-T_{\hat{r}\hat{r}}$ and using
Einstein's field equations, the following relation is obtained

\begin{equation}
\frac{b^{\prime }r-b}{2r^{3}}=-\left[ 2\left( \frac{\dot{\Omega}}{\Omega }
\right) ^{2}-\frac{\ddot{\Omega}}{\Omega }\right] \,.  \label{4boeq}
\end{equation}

In Eq. (\ref{4boeq}) each side depends only on one variable, therefore a
solution is found by separating variables:

\begin{equation}
b(r)=r\left[ 1-\alpha ^{2}(r^{2}-r_{0}^{2})\right] \,,  \label{4solb}
\end{equation}

\begin{equation}
\Omega (t)= \frac{2 \alpha }{C_{1}e^{\alpha t}-C_{2} e^{-\alpha t}},
\label{4solom}
\end{equation}
where $\alpha $ is the separation constant, $C_{1}$ and $C_{2}$ are
constants of integration. If $C_{1}=C_{2}$, $\Omega $ is singular at $t=0$;
therefore for $\alpha >0$ we need to impose $C_{1} > C_{2}$ and if $\alpha
<0 $ it is required that $C_{1} <C_{2}$, otherwise the conformal factor
becomes singular somewhere in the domain of $t$. The conformal function 
$\Omega (t)\rightarrow 0$ as $t\rightarrow \infty $, which reflects a
contracting wormhole solution. It is important to observe that $\Omega
(t)\rightarrow \infty $ as $t\rightarrow t_{0}= \alpha^{-1}{\log (\pm
C_{1}^{-1}\sqrt{C_{1}C_{2}})}$ showing a time singularity that must be
avoided. In the spirit of \cite{AL2} we define the dimensionless parameter 
$R_{0}=\alpha r_{0}$, so that the shape function is given by

\begin{equation}
b(r)=r\left\{ 1-R_{0}^{2}\left[ \left( \frac{r}{r_{0}}\right) ^{2}-1\right]
\right\} \,.  \label{4solb2}
\end{equation}

A fundamental condition to be a reliable wormhole solution is imposed, that 
$b(r)>0$ \cite{wormhole-shell}. Thus, the range of $r$ is $r_{0}<r<a=r_{0} 
\sqrt{1+1/R_{0}^{2}}$; the latter may be arbitrarily large by taking $R_{0}
\rightarrow 0$. If $a\gg r_{0}$, i.e., $R_{0}\simeq r_{0}/a\ll 1$, one may
have an arbitrarily large wormhole.

\subsection{Electromagnetic field equations}

Solving the electromagnetic field Eq. (\ref{em-field}), together with

\begin{equation}
\left( ^{\ast }F^{\mu \nu }\right) _{;\mu }=0\,,  \label{hodge}
\end{equation}
that can be deduced from Bianchi identities, where the asterisk denotes the
Hodge dual \cite{Pleb}, we obtain the restrictions that $%
F_{tr}=-F_{tr}=E(t,r) $, $F_{\theta \phi }=-F_{\phi \theta }=B(\theta )$, $%
L_{F}=L_{F}(t,r)$. Thus the magnetic field is given by

\begin{equation}
B(\theta )=q_{\mathrm{m}}\,\sin \theta \,,  \label{4NB}
\end{equation}
where $q_{\mathrm{m}}$ is a constant related to magnetic charge.

Furthermore, from Eqs. (\ref{4rTott}) and (\ref{4rTohhpp}), we obtain

\begin{equation}  \label{4rTtmh}
\frac{\Omega^2}{8\pi}\left(\frac{b^{\prime}r-3b}{2r^3}\right) 
=\left(1-\frac{b}{r}\right) E^2 L_{F}+ \frac{q_{\mathrm{m}}^2}{r^4}L_{F}.
\end{equation}

The regular solution appears when we set $E=0$. Using Eq. (\ref{4rTtmh}) we
obtain

\begin{equation}
L_{F}=\frac{1}{16\pi q_{\mathrm{m}}^{2}}\Omega ^{2}r(b^{\prime }r-3b)\,,
\label{4NBLF}
\end{equation}
and taking into account Eqs. (\ref{ET}) and (\ref{4rTott}), the lagrangian
is given by

\begin{equation}
L=-\frac{1}{8\pi \Omega ^{2}}\left[ \frac{b^{\prime }}{r^{2}}+3\left( \frac{ 
\dot{\Omega}}{\Omega }\right) ^{2}\right] \,.  \label{4NBL}
\end{equation}

These equations, together with $B=q_{\mathrm{m}}\sin \theta $, $E=0$, $F=q_{%
\mathrm{m}}^{2}/(2\Omega ^{4}r^{4})$ and functions (\ref{4solb}) and (\ref%
{4solom}) give a wormhole solution well behaved at the throat, with finite
fields. This result is in close relationship to the regular magnetic black
holes coupled to nonlinear electrodynamics found in \cite{Bro1}. We remind
that for static solutions with nonlinear electrodynamics, the null energy
condition is not violated, fact that forbids the possibility of wormholes;
however for the evolving wormhole under study this does not apply and
moreover, the weak energy condition is satisfied.

\section{Analysis of the $E=0$ solution}

In what follows we address mainly four aspects of the solution, namely:
flaring out condition and embedded diagram, traversability conditions, the
weak field limit and photon trajectories.

\subsection{Flaring Out Condition and Embedded diagram}

As can be found in \cite{Morris}, \cite{Visser}, an imperative feature to
appear in any wormhole solution is the so called flare out condition once
the embedded diagram is given. The embedding is obtained when one considers
the wormhole metric for an equatorial slice $\theta= \pi /2$ and in a fixed
time $t$, embedding then in a flat three dimensional Euclidean space,
$ds^2= d \overline{z}^2 + d \overline{r}^2+ \overline{r}^2 d \phi^2$. 
 
In our case the flare out condition requires (\cite{AL2}) that

\begin{equation}
\frac{d^{2}\overline{r}\left( \overline{z}\right) }{d\overline{z}^{2}}=\frac{
1}{\Omega \left( t\right) }\frac{d^{2}r\left( z\right) }{dz^{2}}=\frac{1}{
\Omega \left( t\right) }\frac{b-b'r}{2b^{2}}>0  \label{flaoutcond}
\end{equation}
at or near the throat. Using Eqs. (\ref{4solb2}) and (\ref{4solom}) the
flare out condition can be casted as
\begin{equation}
\frac{d^{2}\overline{r}\left( \overline{z}\right) }{d\overline{z}^{2}}=
\frac{r}{2}
\frac{\alpha ( C_{1}\,e^{\alpha t}-C_{2}\,e^{-\alpha t})}
{[1-R_{0}^{2}(r^{2}-r_{0}^{2})]^{2}}>0.
\label{flaoutcondsol}
\end{equation}
 
If $\alpha >0$, then we need to impose $C_{1}e^{\alpha t}>C_{2}e^{-\alpha
t}$, and if $\alpha <0$ and $C_{1}e^{\alpha t}<C_{2}e^{-\alpha t}$, otherwise
the conformal factor becomes negative somewhere along the domain of $t$,
violating the condition $\Omega \left( t\right) >0$. So in general the flare
out condition is fulfilled by the solutions (\ref{4solb2}) and
(\ref{4solom}).
 
Another feature that must be explored for a wormhole solution is the one
related to the form of the function $\overline{z}\left( \overline{r}\right)$
from the embedded diagram \cite{Morris}, \cite{Visser}, this function can be
obtained integrating the relationship

\begin{equation}
\frac{d\overline{z}\left( \overline{r}\right) }{d\overline{r}}=\frac{%
dz\left( r\right) }{dr}=\pm \left[ \frac{r}{b}-1\right] ^{-1/2}.
\label{embeddz}
\end{equation}

This equation can be rewritten using equation (\ref{4solb2}) to obtain

\begin{equation}
\frac{d\overline{z}\left( \overline{r}\right) }{d\overline{r}}=\frac{
dz\left( r\right) }{dr} = \pm \left\{ \frac{1}{\alpha^{2}(
r^{2}-r_{0}^{2}) }-1\right\} ^{1/2}\text{,}  \label{embeddzsol}
\end{equation}
from which we would be able to obtain the solution for $\overline{z}\left( 
\overline{r}\right)$ as
\begin{equation}
\overline{z} \left( \overline{r} \right) = \pm \Omega \left( t \right)
z \left( r \right) = \pm \Omega \left( t \right) \int \left\{
\frac{1}{R_{0}^{2} (r^{2}-r_{0}^{2}) }-1 \right\}^{1/2}dr.  
\label{zint}
\end{equation}
  
\begin{figure}[ht]
\begin{center}
\includegraphics[width=8.4cm,height=6.5cm]{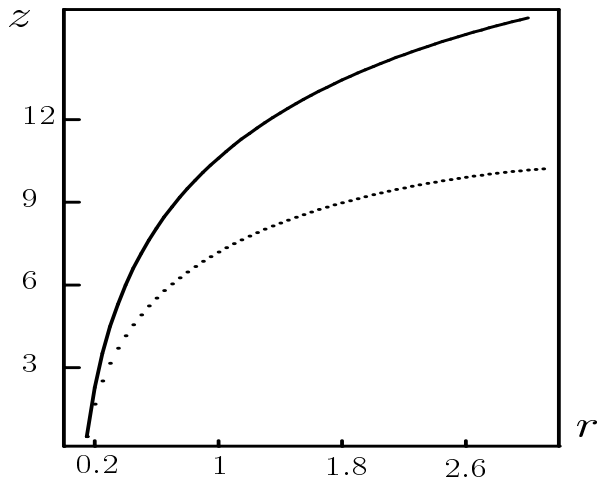}
\end{center}
\caption{ 
The embedded wormhole $z(r)$ for an arbitrary constant time is shown for
distinct values of the parameter $\alpha$. The continuous curve corresponds
to $\alpha=0.2$ while the dotted one is for $\alpha=0.3$}
\label{fig1}
\end{figure}

In Fig. \ref{fig1}, \cite{juanma} it is shown the embedded wormhole $z(r)$
for an arbitrary constant time.  We can observe that the behavior of the
derivative is highly positive near the throat $r_{0}$, as was expected for a
wormhole solution, and as $r$ grows far from the throat, the value of the
derivative $\frac{dz}{dr}$ becomes smaller but positive, indicating that the
tangent lines for $z(r)$ become nearly horizontal. This behavior grant us a
shape for the throat very similar to the typical behavior expected for a
wormhole solution \cite{Morris}, \cite{Visser}.


\subsection{Traversability Conditions}

We explore now the possibility that a human traveler traverses an evolving
wormhole like the one presented here. Leaving aside the study of the
stability of the solution, we delve into the analysis of the tidal
gravitational forces that an infalling radial observer must bear while
traversing. We take as our traversability criteria the magnitude of the tidal
forces that an observer can stand during the trip: it must not exceed the
forces due to Earth gravity, much on the spirit of \cite{Gravitation}. To
simplify our calculations we shall work in a proper reference frame given by

\begin{equation}  \label{travframe}
e_{\hat 0}=\gamma e_{\hat t} \mp \gamma\beta e_{\hat r}\,, \quad e_{\hat
1}=\mp \gamma e_{\hat r} + \gamma\beta e_{\hat t}\,, \quad e_{\hat
2}=e_{\hat \theta}\,, \quad e_{\hat 3}=e_{\hat \phi}\,,
\end{equation}
where $\gamma=\frac{1}{\sqrt{1-\beta^2}}$ and $\beta=v/c$, $v$ is the
velocity of the radial motion \cite{Trobo}. Then the traversability criteria
can be stated as

\begin{equation}  \label{travcritparall}
|R_{\hat 0 \hat 1 \hat 0 \hat 1}|=|C_1\,C_2|\leq\frac{g_{\oplus}}{2 \mathrm{m}}
\approx \frac{1}{(10^{10} \mathrm{cm})^2}\,,
\end{equation}

\begin{equation}  \label{travcritangul}
|R_{\hat 0 \hat 2 \hat 0 \hat 2}|=|R_{\hat 0 \hat 3 \hat 0 \hat 3}|=\gamma^2
|R_{\hat \theta \hat t \hat \theta \hat t}|+\gamma^2\beta^2 |R_{\hat \theta
\hat r \hat \theta \hat r}|=|C_1\,C_2|\leq\frac{g_{\oplus}}{2 \mathrm{m}}
\approx\frac{ 1}{(10^{10}\mathrm{cm})^2}\,,
\end{equation}
where $g_{\oplus}=9.81$m/s$^2$ is Earth gravity; these conditions can be
fulfilled with the appropriate selection of the values for $C_1$ and $C_2$,
rendering this solution in a traversable one. We point out the fact that Eq.
(\ref{travcritangul}) does not depend on the traverse velocity $v$, however
it is not in contradiction with the results given in \cite{Visser}, Chap.
13.1.1., related to the dependence of tidal forces on the traveler velocity,
due to the fact that the condition $R_{\hat\theta\hat t\hat\theta\hat
t}=-R_{\hat\theta\hat r\hat\theta\hat r}$ does not lead to the presence of a
horizon, instead, it take us to equation (\ref{4boeq}) from which the
solution is obtained.

Since in the studied case the redshift has been set equal to zero, the
radial tides are zero, then if some danger exists for the traveler it is in
the nonzero accelerations while traversing the wormhole. In what follows we
show that there is not tidally induced shear.

The $Q_{IJ}$ matrix that describes tidal forces is defined by

\begin{equation}
Q_{IJ} =-R_{\mu \alpha \nu
\beta}(\eta_{I})^{\mu}V^{\alpha}(\eta_{J})^{\nu}V^{\beta},
\end{equation}
where $\eta_{J}, J=1,2,3$ is an orthonormal triad orthogonal to the
four-velocity $V^{\alpha}$. Let us consider the traveler's motion is non
radial with a velocity vector given by

\begin{equation}
V^{\hat{\mu}}=(\gamma, \gamma \beta \cos \psi, 0, \gamma \beta \sin \psi ).
\label{nonradialveloc}
\end{equation}

In the case we are analyzing, the tidal forces are described with the
entrances

\begin{equation}
Q_{11}=-R_{\hat{t}\hat{r}\hat{t}\hat{r}}\cos ^{2}\psi -R_{\hat{t}\hat{\phi} 
\hat{t}\hat{\phi}}\sin ^{2}\psi =C_{1}C_{2}\,,  \label{Q11}
\end{equation}
\begin{equation}
Q_{22}=-\gamma ^{2}(R_{\hat{t}\hat{r}\hat{t}\hat{r}}\sin ^{2}\psi +R_{\hat{t}
\hat{\phi}\hat{t}\hat{\phi}}\cos ^{2}\psi +\beta ^{2}R_{\hat{r}\hat{\phi} 
\hat{r}\hat{\phi}})=C_{1}C_{2}\,,  \label{Q22}
\end{equation}
\begin{eqnarray}
Q_{33} &=&-\gamma ^{2}(R_{\hat{t}\hat{\phi}\hat{t}\hat{\phi}}+\beta ^{2}R_{ 
\hat{r}\hat{\phi}\hat{r}\hat{\phi}}\cos ^{2}\psi +\beta ^{2}R_{\hat{\theta} 
\hat{\phi}\hat{\theta}\hat{\phi}}\sin ^{2}\psi )  \label{Q33} \\
&=&-\gamma ^{2}[-C_{1}C_{2}+C_{1}C_{2}\beta ^{2}\cos ^{2}\psi +\frac{\beta
^{2}\sin ^{2}\psi }{4}[(C_{1}e^{\alpha t}+C_{2}e^{-\alpha t})^{2}]  \nonumber
\\
&&+\frac{1}{\alpha ^{2}r^{2}}(1-\alpha ^{2}(r^{2}-r_{\mathrm{0}
}^{2}))(C_{1}e^{\alpha t}-C_{2}e^{-\alpha t})^{2}]\}\,,  \nonumber
\end{eqnarray}
\begin{equation}
Q_{12}=-\gamma (R_{\hat{t}\hat{r}\hat{t}\hat{r}}-R_{\hat{t}\hat{\phi}\hat{t} 
\hat{\phi}})\sin \psi \,\cos \psi =0\,,  \label{Q12}
\end{equation}
and 
\begin{equation}
Q_{13}=Q_{23}=0\,.  \label{Q13-23}
\end{equation}

Therefore, the tidal forces do not diverge, furthermore, since $Q_{12}=0$,
there is no presence of tidally induced shear, this means that the traveler
will pass safe through the wormhole as far as tidal forces concern.

\subsection{Maxwellian Limit.}

It is important that any nonlinear theory recovers the form and results of
the corresponding linear theory, nonlinear electrodynamics is not an
exception \cite{Pleb}. However there are cases where the Maxwellian limit is
not recovered \cite{Bardeen-Beato}.

To carry on the analysis of the Maxwell limit, it is useful to rewrite the
main functions $F$, $L$ and $L_{F}$ in terms of the coordinates $r$ and $t$,

\begin{equation}
F=\frac{q_{\mathrm{m}}^{2}}{32\alpha ^{4}}\left( \frac{C_{1}e^{\alpha
t}-C_{2}e^{-\alpha t}}{r}\right) ^{4}\,,  \label{Foftandr}
\end{equation}

\begin{equation}
L=-\frac{1}{32\pi \alpha ^{2}}\left\{ [1-\alpha ^{2}(3r^{2}-r_{\mathrm{0}
}^{2})][\frac{C_{1}e^{\alpha t}-C_{2}e^{-\alpha t}}{r}]^{2}+3\alpha
^{2}(C_{1}e^{\alpha t}+C_{2}e^{-\alpha t})^{2}\right\} \,,  \label{Loftandr}
\end{equation}

\begin{equation}
L_{F}=-\frac{\alpha ^{2}(1+\alpha ^{2}r_{\mathrm{0}}^{2})}{2\pi q_{\mathrm{m}
}^{2}}[\frac{r}{C_{1}e^{\alpha t}-C_{2}e^{-\alpha t}}]^{2}.
\label{LFoftandr}
\end{equation}
 
The result of the analysis of the solution is the absence of a complete
Maxwellian limit, i.e., when $r\rightarrow \infty $ we get that $F
\rightarrow 0$, $L \rightarrow -\frac{3C_{1}C_{2}}{8\pi }$ and $L_{F}
\rightarrow \infty $. The fact that the L lagrangian $L$ does not go as $F$,
and instead goes to a constant value shows that in this limit the spacetime
is not asymptotically flat due to a constant remanent energy. Another
impressive result is the corresponding to $L_{F}$ which goes to infinity,
showing a similar behavior to an ideal conductor \cite{Dymnikova-Burinskii}.
 
Alternatively, as the solution is only valid for the region $r_{0}<r<a=r_{0}%
\sqrt{1+1/\beta ^{2}}$, and the Maxwellian limit would be attained as $r
\rightarrow \infty $, the requirement of the presence of a region with
Maxwellian limit losses strength. Furthermore, it does make sense the absence
of such a limit since for weak fields the energy conditions are always
fulfilled, a fact that is contrary to the existence of wormholes.

\subsection{Light rays in the NLED effective geometry}

In nonlinear electrodynamics photons do not propagate along null geodesics of
the background geometry, instead they propagate along null geodesics of an
effective geometry which depends on the nonlinear theory \cite{Pleb},
\cite{Novello}. The discontinuities of the field propagate obeying the
equation for the characteristic surfaces $S$. For a curved spacetime the
equation for the characteristic surfaces is

\begin{equation}
g^{\mu \nu}S_{, \mu}S_{, \nu}=0
\end{equation}

And when nonlinear electrodynamics is involved, the corresponding ``eikonal"
equation for the propagation vectors $k^{\mu}$ is

\begin{equation}
\left( L_{F} g^{\mu \nu}- 4 L_{FF}F^{\mu \alpha}F_{\alpha}^{\nu}
\right)k_{,\mu}k_{,\nu}=g_{eff}^{{\mu}{\nu}}k_{, \mu}k_{, \nu}=0
\end{equation}

In the orthonormal tetrad, the expression of $g_{eff}^{\hat{\mu}\hat{\nu}}$
in terms of the electromagnetic field tensor $T^{\hat{\mu}\hat{\nu}}$ also
shows clearly that the modification in the trajectories is due to nonlinear
electromagnetic field:

\begin{equation}
g_{eff}^{\hat{\mu}\hat{\nu}}=\left( L_{F}+\frac{LL_{FF}}{L_{F}}\right) 
\eta^{\hat{\mu}\hat{\nu}}+\frac{L_{FF}}{L_{F}}T^{\hat{\mu}\hat{\nu}},  \label{1}
\end{equation}
given in orthonormal coordinates $\hat{\mu}\hat{\nu},$ with $\eta^{\hat{\mu} 
\hat{\nu}}= \mathrm{diag} [-1,1,1,1]$. For the case of the evolving wormhole
with nonlinear electrodynamics (\ref{4dysme}) with (\ref{4solb}) and 
(\ref{4solom}), the effective geometry is given by

\begin{eqnarray}
g_{eff}^{\hat{t}\hat{t}} &=&-\left( L_{F}+2L\frac{L_{FF}}{L_{F}}\right)
=-g_{eff}^{\hat{r}\hat{r}},  \label{2} \\
g_{eff}^{\hat{\theta}\hat{\theta}} &=&L_{F}+2L\frac{L_{FF}}{L_{F}}
-2FL_{FF}=g_{eff}^{\hat{\varphi}\hat{\varphi}},  \label{4}
\end{eqnarray}
where we have taken into account the energy-momentum tensor given by

\[
\mathrm{diag} (T^{\hat{t}\hat{t}},T^{\hat{r}\hat{r}},T^{\hat{\theta}
\hat{\theta}}, T^{\hat{\phi}\hat{\phi}}) 
\]
with 
\begin{eqnarray}
T^{\hat{t}\hat{t}} &=&-L =-T^{\hat{r}\hat{r}},  \label{5} \\
T^{\hat{\theta}\hat{\theta}} &=&L-2FL_{F}=T^{\hat{\phi}\hat{\phi}},
\label{6}
\end{eqnarray}
that are Eqs. (18)-(20) in \cite{AL2}. Moreover, the expression for the
invariant $F$, if $E=0$ is

\begin{equation}
F=\frac{1}{2}\frac{q_{m}^{2}}{\Omega ^{4}r^{4}}.  \label{7}
\end{equation}

Now, in the particular case that $b(r)=r\left[ 1-\alpha ^{2}\left(
r^{2}-r_{0}^{2}\right) \right],$ we have the relationship 
\begin{equation}
\frac{L_{FF}}{L_{F}}=-\frac{1}{F},  \label{8}
\end{equation}
therefore Eqs. (\ref{2})-(\ref{4}) amount to

\begin{eqnarray}
g_{eff}^{\hat{t}\hat{t}} &=&-\left( L_{F}-2\frac{L}{F}\right)
=-g_{eff}^{\hat{r}\hat{r}} ,  \label{9} \\
g_{eff}^{\hat{\theta}\hat{\theta}} &=&3L_{F}-2\frac{L}{F}=g_{eff}^{\hat{%
\varphi}\hat{\varphi}}.  \label{11}
\end{eqnarray}

In the effective metric the geodesic equations are

\begin{eqnarray}
\frac{d^{2}t}{d\tau ^{2}}+\left( \frac{\dot{L}_{F}F^{2}-2\dot{L}F+2L\dot{F}}{
2F(FL_{F}-2L)}\right) \left( \left( \frac{dt}{d\tau }\right) ^{2}-\left(
\frac{dr}{d\tau }\right) ^{2}\right) +\left( \frac{2\dot{L}F-2L\dot{F}-3\dot{
L}_{F}F^{2}}{2F(FL_{F}-2L)}\right) A^{2} &=&0 \label{G1} \\
\frac{d^{2}r}{d\tau ^{2}}-\left( \frac{L_{F}^{\prime 2}-2L^{\prime
}F+2LF^{\prime }}{2F(FL_{F}-2L)}\right) \left( \left( \frac{dt}{d\tau }
\right) ^{2}-\left( \frac{dr}{d\tau }\right) ^{2}\right) +\left(
\frac{-2L^{\prime }F+2LF^{\prime }+3L_{F}^{\prime
}F^{2}}{2F(FL_{F}-2L)}\right) A^{2} &=&0
\label{G2} \\
\left( L_{F}-\frac{2L}{F}\right) \left( -\left( \frac{dt}{d\tau }\right)
^{2}+\left( \frac{dr}{d\tau }\right) ^{2}\right) +\left( 3L_{F}-2\frac{L}{F}
\right) A^{2} &=&\delta \label{G3} \\ \left( \left( \frac{d\theta }{d\tau
}\right) ^{2}+\left( \frac{d\varphi }{ d\tau }\right) ^{2}\right) &=&A^{2},
\end{eqnarray}
where $\tau $ is the affine parameter that generates the geodesic
trajectories $r(\tau ),t(\tau ),\theta (\tau )$, $\varphi (\tau )$. $A$ is a
conserved quantity related to the existence of Killing vectors. The Eq. 
(\ref{G3}) is derived from the line element and $\delta =1,0,-1$ for timelike,
null and spacelike geodesics, respectively.
 
Substituting $\left( (\frac{dr}{d\tau})^{2}-(\frac{dt}{d\tau})^{2}\right)$
from Eq.(\ref{G3}) in Eqs. (\ref{G1}) and (\ref{G2}), for $\delta=0$ it is
obtained

\begin{eqnarray}
\frac{d^2t}{d\tau^2}+\frac{2\left( LL_{F} \dot{F}+L \dot{L}_{F}F-L_{F} \dot{L%
}F\right)}{\left( FL_{F}-2L\right)^{2}}A^2 &=&0, \\
\frac{d^2r}{d\tau^2}+\frac{2\left( LL_{F}F^{\prime }+LL_{F}^{\prime
}F-L_{F}L^{\prime }F\right)}{\left( FL_{F}-2L\right)^{2}} A^2&=&0.
\end{eqnarray}

The constant of motion $A$ makes easier the integration of the geodesic
equations. Considering the expressions for $L_{F}, L, F$ in terms of $r$ and 
$t$, straightforward algebra leads to the expressions

\begin{eqnarray}
0&=&\frac{d^2t}{d\tau^2}+ \frac{A^2}{2} \frac{df(r,t)}{dt},  \nonumber \\
0&=&\frac{d^2r}{d\tau^2}+ \frac{A^2}{2} \frac{dg(r,t)}{dr},  \label{geod1}
\end{eqnarray}
with 
\begin{eqnarray}
f(r,t)&=& -\frac{(1+\alpha^2 r_0^2)}{6r^2} \left[ \frac{(1+\alpha^2 r_0^2)}
{4r^2}- \alpha^2 + \left( \frac{\dot \Omega}{\Omega}\right)^2 \right]^{-1}, 
\nonumber \\
g(r,t)&=& \frac{2}{3} \left[\alpha^2 -\left( \frac{\dot\Omega}{\Omega}
\right)^2 \right] \left[ \frac{(1+\alpha^2 r_0^2)}{4r^2}- \alpha^2 + \left( 
\frac{\dot \Omega}{\Omega}\right)^2 \right]^{-1}.  \label{geod2}
\end{eqnarray}
 
Eqs. (\ref{geod1}) can be integrated by multiplying the first equation by 
$2(dt)/(d\tau )$ and the second by $2(dr)/(d\tau )$, so they acquire
the form of

\begin{eqnarray}
\frac{d}{d\tau}\left[\left( \frac{dt}{d \tau}\right)^2+A^2 f(r,t)\right]&=&0,
\\
\frac{d}{d\tau}\left[\left( \frac{dr}{d \tau}\right)^2+ A^2g(r,t)\right]&=&0,
\end{eqnarray}
that allows one to obtain the trajectories $r(t)$ of photons as

\begin{equation}
\frac{dr}{dt}= \sqrt{\frac{K_2-g(r,t)}{K_1-f(r,t)}},
\end{equation}
where $K_1$ and $K_2$ are constants. Substituting $f(r,t)$ and $g(r,t)$ from
(\ref{geod2}) and simplifying we get

\begin{equation}
\frac{dr}{dt}= \sqrt{\frac{(1+\alpha^2 r_0^2)K_2(Be^{2 \alpha t}-1)^2+ 16
\alpha^2 r^2B(K_2+2/3)e^{2 \alpha t}} {(1+\alpha^2 r_0^2)(K_1-2/3)(Be^{2
\alpha t}-1)^2+ 16 \alpha^2 r^2BK_1e^{2 \alpha t}}},
\end{equation}
where $B=C_1/C_2$, $B<1$ for the square root be well defined; $C_1$ and $C_2$
can be adjusted to fulfil the traversability conditions as well.
 
\begin{figure}[ht]
\begin{center}
\includegraphics[width=8.4cm,height=6.5cm]{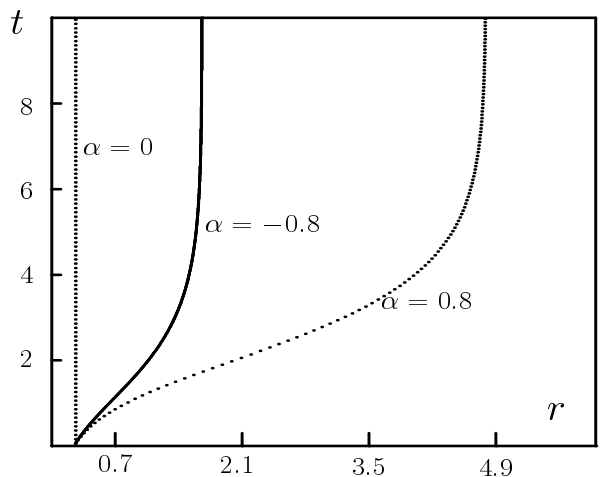}
\end{center}
\caption{Null geodesics $r(t)$ in the effective geometry of the nonlinear
electromagnetic field. In these plots $K_1=K_2=0$ and the static case is for 
$\protect\alpha=0$. The non-flatness of the spacetime is apparent, as the
geodesics do not tend to infinity for large times}
\label{fig2}
\end{figure}

\begin{figure}[ht]
\begin{center}
\includegraphics[width=8.4cm,height=6.5cm]{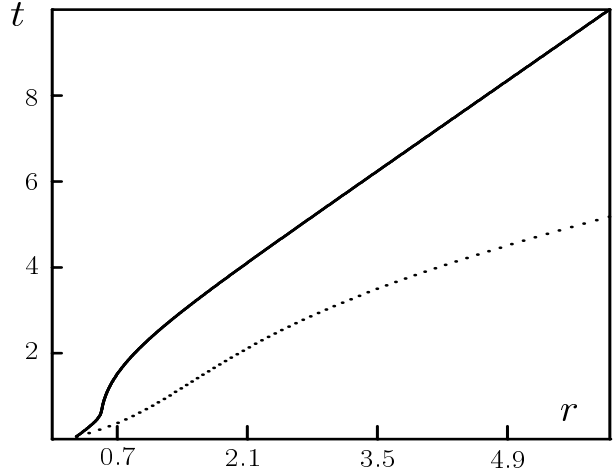}
\end{center}
\caption{Trajectories of photons $r(t)$ in the effective geometry. The
continuous curve corresponds to $K_1=1.8, K_2=0.5$, while the dotted one is
for $K_1=0.5, K_2=-0.66$, in both $\protect\alpha=0.8$. The nonvanishing of
the constants of movement enhances the action of the magnetic field and
smooths the non-flatness of the spacetime}
\label{fig3}
\end{figure}

The graphics \cite{juanma} of the trajectories $r(t)$ are shown in Fig.\ref
{fig2} for different values of $\alpha$, with $K_1=K_2=0$, showing the
influence of the magnetic field variation. Fig.\ref{fig3} shows $r(t)$ when
$K_1$ and $K_2$ are not zero. It can be observed in both plots the
nonvanishing throat, as well as the non-flatness of the spacetime (associated
to the non-Maxwellian limit). The effective spacetime as ``seen" by photons
does not carry as much restrictions as the mere existence of the wormhole.

\section{Evolving wormhole solution with two invariants $F$ and $G$}

The previous analyzed wormhole corresponds to a lagrangian that depends only
on one of the two electromagnetic invariants, $F=F^{\mu \nu }F_{\mu \nu }/4.$
On the search for a more general solution we explore in this
section the case of a lagrangian depending in nonlinear way on both
invariants.

Let us consider the action of $(3+1)-$dimensional general relativity coupled
to nonlinear electrodynamics given by

\begin{equation}
S=\int\sqrt{-g}\left( \frac{R}{16\pi}+L(F,G)\right) d^{4}x,  \label{S}
\end{equation}
where $R$ is the Ricci scalar. $L(F,G)$ is a gauge-invariant electromagnetic
lagrangian depending on the invariants $F=\frac{1}{4}F^{\mu\nu}F_{\mu\nu} $
and $G=\frac{1}{4}F^{\mu\nu}\check{F}_{\mu\nu}$ where $F_{\mu\nu}$ is the
electromagnetic tensor and $\check{F}_{\mu\nu}=\frac{1}{2}
\epsilon_{\mu\nu\alpha\beta}F^{\alpha\beta}$ its dual.

Varying the action with respect to the gravitational field provides the
Einstein field equations $G_{\mu\nu}=8\pi T_{\mu\nu},$ with the
stress-energy tensor given by

\begin{equation}
T_{\mu\nu}=g_{\mu\nu}L-F_{\mu\delta}F_{\nu}^{\quad\delta}L_{F}-F_{\mu\delta
} \check{F}_{\nu}^{\quad\delta}L_{G},  \label{Tmunu}
\end{equation}
where $L_{F}=\frac{\partial L}{\partial F}$ and $L_{G}=\frac{\partial L}{
\partial G}$ are functions of $r$ and $t$ only, in the same form $L=L(t,r).$
The variation with respect to the electromagnetic potential $A_{\mu},$
yields the electromagnetic field equations

\begin{equation}
p_{\quad;\nu}^{\mu\nu}\equiv\left( F^{\mu\nu}L_{F}+\check{F}
^{\mu\nu}L_{G}\right) _{;\nu}=0.  \label{EFE}
\end{equation}

We shall consider a spacetime metric representing a dynamic spherically
symmetric $(3+1)-$dimensional wormhole given by (\ref{4dysme}). The non-zero
Einstein tensor components are the same that in Eq. (\ref{ET}).
 
The electromagnetic tensor, compatible with the symmetries of the geometry,
is given by (\ref{4em-tensor}). The $F$ invariant is given by Eq.
(\ref{invF}) while the invariant $G$ is 
\[G=\frac{\sqrt{1-\frac{b(r)}{r}}EB}{\Omega ^{4}r^{2}\sin (\theta )}.\]

The components of the stress-energy tensor, Eq. (\ref{Tmunu}), in the
orthonormal frame take the following form

\begin{eqnarray}
T_{\hat{t}\hat{t}} & =&-T_{\hat{r}\hat{r}}=-L-\frac{E^{2}L_{F}\left( 1-\frac{%
b }{r}\right) }{\Omega^{4}}+\frac{EBL_{G}\sqrt{1-\frac{b}{r}}}{
\Omega^{4}r^{2}\sin(\theta)}  \nonumber \\
& =&-L-\frac{E^{2}L_{F}\left( 1-\frac{b}{r}\right) }{\Omega^{4}}+GL_{G}, 
\nonumber \\
T_{\hat{\theta}\hat{\theta}} & =&T_{\hat{\varphi}\hat{\varphi}}=L-\frac {
EBL_{G}\sqrt{1-\frac{b}{r}}}{\Omega^{4}r^{2}\sin(\theta)}-\frac{B^{2}L_{F}}{
\Omega^{4}r^{4}\sin^{2}(\theta)}  \nonumber \\
& =&L-GL_{G}-\frac{B^{2}L_{F}}{\Omega^{4}r^{4}\sin^{2}(\theta)},  \nonumber
\\
T_{\hat{t}\hat{\imath}} & =&T_{\hat{\imath}\hat{\jmath}}=0\text{ (with }
i\neq j\text{).}
\end{eqnarray}

As in \cite{AL2}, $T_{\hat{t}\hat{t}}=-T_{\hat{r}\hat{r}},$ so the solution
to $G_{\hat{t}\hat{t}}=-G_{\hat{r}\hat{r}}$ can be solved separating
variables like in (\ref{4boeq}): 
\begin{eqnarray}
b(r)& =&r\left[ 1-\alpha ^{2}\left( r^{2}-r_{0}^{2}\right) \right] , \\
\Omega (t)& =&\frac{2\alpha }{C_{1}e^{\alpha t}-C_{2}e^{-\alpha t}},
\end{eqnarray}
where $\alpha $ is a constant, and $C_{1}$ and $C_{2}$ are constants of
integration.

The electromagnetic field Eqs. (\ref{EFE}) take the form 
\begin{eqnarray}
\sqrt{1-\frac{b}{r}}\partial_{t}(EL_{F})-\frac{B\left(
\partial_{t}L_{G}\right) }{r^{2}\sin(\theta)} & =&0,  \label{EFE1} \\
\partial_{r}\left( \sqrt{1-\frac{b}{r}}r^{2}EL_{F}\right) -\frac{B\left(
\partial_{r}L_{G}\right) }{\sin(\theta)} & =&0,  \label{EFE2} \\
\frac{\left[ B\cos(\theta)-\partial_{\theta}B\sin(\theta)\right] L_{F}}{
r^{4}\Omega(t)^{4}\sin(\theta)^{3}} & =&0.  \label{EFE3}
\end{eqnarray}

Eq. (\ref{EFE1}) can be solved for $EL_{F}:$ 
\begin{equation}
EL_{F}=\frac{BL_{G}}{r^2 \sqrt{1-\frac{b}{r}} \sin(\theta)}+F_{1}(r).
\label{S1}
\end{equation}
Now, Eq. (\ref{EFE2}), when we take into account Eq. (\ref{S1}) give us the
expression for $F_{1}(r)$
 
\[ F_{1}(r)=\frac{C_{E}}{ r^2 \sqrt{1-\frac{b}{r}}},\qquad
C_{E}=\mathrm{cte.} \] 
Then (\ref{S1}) takes the form

\begin{equation}
EL_{F}=\frac{1}{ r^2 \sqrt{1-\frac{b}{r}}}\left[ \frac{BL_{G}}{\sin
(\theta )
}+C_{E}\right] ,  \label{ELF}
\end{equation}
From this last relation we verify that $EL_{F}$ is singular at the throat.
Integrating Eq. (\ref{EFE3}), we obtain $B=q_{m}\sin (\theta ).$ Of course
when $L_{G}=0,$ we recover the expression obtained in \cite{AL2} for
$EL_{F}$.

Therefore, when there is dependence on both electromagnetic invariants, $F$
and $G,$ we recover the solution in \cite{AL2}, except for an additional term
$\kappa G,$

\[ L=L(F)+\kappa G \]
 
This term breaks the duality rotation symmetry of the electromagnetic
field, present for instance in nonlinear electrodynamics of the
Born-Infeld type.

We conclude that geometry (4) does not allow nonlinear electromagnetic matter
related to a lagrangian $L(F,G)$. There is open still the question if
assuming a less symmetrical spacetime it would admit such kind of matter.

\section{Final Remarks}

In this article we have analyzed the solution representing an evolving
wormhole coupled to nonlinear electrodynamics given by Arellano and Lobo in
\cite{AL2}. We presented the integration of the geodesic equations for light
rays and the corresponding graphics that show the influence of the nonlinear
field; the traversability conditions as well as the weak field limit were
analyzed.

In the conclusions about traversability, we obtained that the tidal
gravitational forces are constant in the radial movement, and they can be
adjusted to allow a safe human passage.

For the Maxwellian limit we have found that when $r\rightarrow \infty$ the
spacetime is asymptotically flat except for a remanent energy, a possible
interpretation can be the presence of a magnetic perfect fluid that can be
obtained from the stress energy tensor taking that limit. We must stress that
the wormhole solution is only valid in a certain region that, even if it can
be extended to be large, leaves outside the limit $r\rightarrow \infty $.
However, the allowed trajectories for photons show that the NLED effective
geometry is less restrictive than the one for the wormhole.

When the dependence on both invariants for the lagrangian is considered, it
turns out that the electric-magnetic duality symmetry is lost.

\section*{Acknowledgments}

A. V. B. A. acknowledges a Ph. D. grant by Conacyt-Mexico. N. B.
acknowledges partial support by Conacyt-Mexico, project 49182-F. R. G-S. was
partially financed by Conacyt-Mexico, project 49924-J.

\end{document}